\documentclass[reprint,amssymb,amsmath,aip,rsi]{revtex4-1}
\usepackage{bm}
\usepackage{graphicx}
\usepackage[colorlinks=true,linkcolor=blue]{hyperref}
\expandafter\ifx\csname package@font\endcsname\relax\else
 \expandafter\expandafter
 \expandafter\usepackage
 \expandafter\expandafter
 \expandafter{\csname package@font\endcsname}
\fi
\begin{document}
\title{Ultrasound measurement technique for the single-turn-coil magnets}
\author{T. Nomura}
\email{t.nomura@issp.u-tokyo.ac.jp}
\affiliation{Institute for Solid State Physics, University of Tokyo, Kashiwa, Chiba 277-8581, Japan}
\author{A. Hauspurg}
\affiliation{Hochfeld-Magnetlabor Dresden (HLD-EMFL) and W\"urzburg-Dresden Cluster of Excellence ct.qmat, Helmholtz-Zentrum Dresden-Rossendorf, 01328 Dresden, Germany}
\affiliation{Institut f\"ur Festk\"orper- und Materialphysik, Technische Universit\"at Dresden, 01062 Dresden, Germany}
\author{D. I. Gorbunov}
\affiliation{Hochfeld-Magnetlabor Dresden (HLD-EMFL) and W\"urzburg-Dresden Cluster of Excellence ct.qmat, Helmholtz-Zentrum Dresden-Rossendorf, 01328 Dresden, Germany}
\author{A. Miyata}
\affiliation{Hochfeld-Magnetlabor Dresden (HLD-EMFL) and W\"urzburg-Dresden Cluster of Excellence ct.qmat, Helmholtz-Zentrum Dresden-Rossendorf, 01328 Dresden, Germany}
\author{E.~Schulze}
\affiliation{Hochfeld-Magnetlabor Dresden (HLD-EMFL) and W\"urzburg-Dresden Cluster of Excellence ct.qmat, Helmholtz-Zentrum Dresden-Rossendorf, 01328 Dresden, Germany}
\affiliation{Institut f\"ur Festk\"orper- und Materialphysik, Technische Universit\"at Dresden, 01062 Dresden, Germany}
\author{S. A. Zvyagin}
\affiliation{Hochfeld-Magnetlabor Dresden (HLD-EMFL) and W\"urzburg-Dresden Cluster of Excellence ct.qmat, Helmholtz-Zentrum Dresden-Rossendorf, 01328 Dresden, Germany}
\author{V. Tsurkan}
\affiliation{Experimental Physics V, Center for Electronic Correlations and Magnetism, Institute of Physics, University of Augsburg, 86159, Augsburg, Germany}
\affiliation{Institute of Applied Physics, MD-2028, Chisinau, Republic of Moldova}
\author{Y. H. Matsuda}
\affiliation{Institute for Solid State Physics, University of Tokyo, Kashiwa, Chiba 277-8581, Japan}
\author{Y. Kohama}
\affiliation{Institute for Solid State Physics, University of Tokyo, Kashiwa, Chiba 277-8581, Japan}
\author{S. Zherlitsyn}
\affiliation{Hochfeld-Magnetlabor Dresden (HLD-EMFL) and W\"urzburg-Dresden Cluster of Excellence ct.qmat, Helmholtz-Zentrum Dresden-Rossendorf, 01328 Dresden, Germany}
\date{\today}

\begin{abstract}
Ultrasound is a powerful means to study numerous phenomena of condensed-matter physics as acoustic waves couple strongly to structural, magnetic, orbital, and charge degrees of freedom.
In this paper, we present such technique combined with single-turn coils (STC) which generate magnetic fields beyond 100~T with the typical pulse duration of 6~$\mu$s.
As a benchmark of this technique, the ultrasound results for MnCr$_2$S$_4$, Cu$_6$[Si$_6$O$_{18}$] $\cdot$ 6H$_2$O, and liquid oxygen are shown.
The resolution for the relative sound-velocity change in the STC is estimated as $\Delta v/v \sim10^{-3}$, which is sufficient to study various field-induced phase transitions and critical phenomena.
\end{abstract}
\maketitle

\section{Introduction}
Ultrasound technique is a powerful tool in condensed-matter physics to study various materials, ranging from liquids to solids, and from metals to insulators \cite{Luthi05,Bhatia}.
The sound velocity $v$ is related to the elastic constant $c=\rho v^2$, where $\rho$ is the mass density.
The elastic constant is the second-order derivative of the free energy with respect to strain.
Therefore, the ultrasound technique is known to be very sensitive to numerous phase transitions (both structural and magnetic) which are not always detected in specific heat or magnetic susceptibility experiments \cite{11PRL_Tsurkan,13PRL_Tsurkan,15Zherlitsyn,17SA_Tsurkan,19PRR_Gen}.

The ultrasound technique has been widely employed to study field-induced phenomena in solids.
The target materials include frustrated magnets \cite{11PRL_Tsurkan,13PRL_Tsurkan,15Zherlitsyn,Wolf01,17SA_Tsurkan,19PRR_Gen}, rare-earth inter-metallic compounds \cite{11PRB_Andreev,14PRB_Gorbunov,18PRB_Gorbunov,20PRB_Gorbunov}, heavy-fermion systems \cite{99PRB_Zherlitsyn,20PRB_Kurihara}, Weyl semimetals \cite{Ramshaw18,20PRBSchindler} and so on.
In those studies, non-destructive pulse-field magnets are used up to the fields of 90~T with typical pulse durations of 10 -- 200 ms.
Within this time scale, the standard pulse-echo (PE) technique with the repetition rate of $\sim50$~kHz can be employed \cite{01Physica_Wolf,Zherlitsyn14}.
For higher magnetic fields, only destructive pulsed-field magnets with a few $\mu$s pulse duration can be currently used \cite{Nakao85,Miura03,Portugall99,Portugall13,Mielke06}.
However, the $\mu$s time scale is too short to use the PE method, making ultrasound measurements in magnetic field above 100 T a challenge.

In this paper, we present a new approach which combines the ultrasound method and ultrahigh magnetic fields generated by a single-turn coil (STC).
The performance of this technique is illustrated using several magnetic materials exhibiting field-induced phase transitions.
The new experimental approach can be extended up to the field of 200 T.

This paper is organized as follows.
In Sec. II, we discuss the experimental challenges for the ultrasound measurements in the STC.
Afterwards, we introduce our experimental strategy, the continuous-wave (CW) excitation, and compare with the standard PE method.
In Sec. III, we present and discuss the obtained experimental results.
In Sec. IV, conclusive remarks are given.

\section{Experiment}
\subsection{Experimental challenges}

\begin{figure}[tb]
\centering
\includegraphics[width=0.98\linewidth]{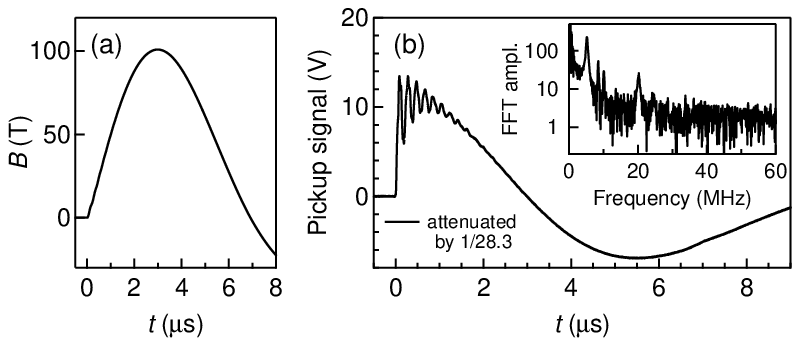}
\caption{
(a) Typical magnetic field versus time waveform of the STC.
(b) Pickup-coil voltage, which is proportional to $dB/dt$.
The signal was attenuated by the voltage attenuator with the ratio of 1/28.3.
The Fourier-transform spectra is shown in the inset.
} 
\label{field}
\end{figure}

The goal of this work is to combine the STC pulsed magnets and ultrasound measurements.
The STC is a so-called semi-destructive way to generate magnetic field, where the STC explodes outwards, but the whole installation inside of the coil (cryostat tail, sample, and so on) usually survives \cite{Nakao85,Miura03}.
The highest magnetic field for physical experiments is about 200~T with the inner magnet bore of 8~mm.
Physical property measurements with the STC are generally challenging because of (i) the very short pulse duration (a few $\mu$s), (ii) significant electromagnetic noise, and (iii) shock as a result of the coil explosion.
Let us discuss these challenges in relation to ultrasound measurements.
Figure \ref{field} shows a typical magnetic-field waveform $B(t)$ and detected $dB/dt(t)$ voltage of the STC.
i) First, the total field-duration time is $\sim 6$ $\mu$s (corresponding to $\sim100$ kHz), and the peak duration time (above 90~\% of the maximum field) is only 1 $\mu$s.
Compared to the typical repetition rate of the ultrasound PE technique (50~kHz) in non-destructive pulsed magnets, the field-sweep rate of the STC is too high.
ii) Second, as evident from the $dB/dt(t)$ dependence, the signal from 0 to 1 $\mu$s is strongly disturbed by the discharge noise.
Because of the spark-gap switch, huge electromagnetic noise is produced at the beginning of the pulse.
Here, the oscillatory noise from 1 to 25 MHz is also present, which comes from the impedance mismatch of the STC system.
Besides, if there is an open loop in an electrical circuit, pickup voltage of the order of 100 V can easily damage the measurement instruments.
Therefore, the intensity of the useful signal has to be high enough and the noise from the STC system has to be suppressed.
iii) Third, when the STC explodes, a shock wave is generated, which can mechanically damage the sample installation.
Even though the explosion occurs outwards, the experimental configuration inside the coil has to be robust against the mechanical disturbance.

The biggest obstacle for the ultrasound measurement is (i) the very short magnetic-field duration.
In the STC magnet, the measurement sampling rate has to be higher than a few MHz.
Besides, typical sound velocity in solids is $v=2$ -- 7~km/s.
If the sample length is a few millimeters, the propagation time is of the order of 1 $\mu$s.
This time delay is comparable with the field duration of the STC.
Thus, the sound-velocity measurement cannot be done at the quasi-static field conditions, but is rather in a transient regime.
In this case, even if the sound velocity changes discontinuously, the detected sound velocity is averaged during the sound propagation time $\tau$.

\subsection{Continuous-wave excitation technique}

\begin{figure}[tb]
\centering
\includegraphics[width=0.95\linewidth]{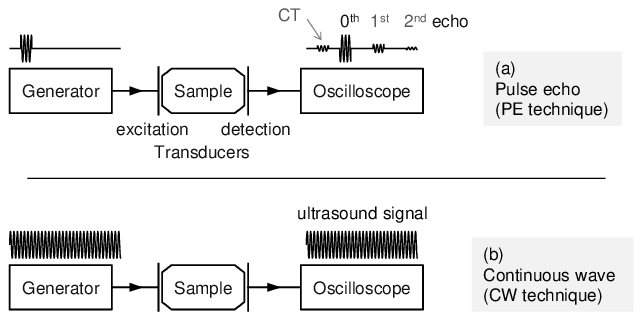}
\caption{
Schematics of the (a) pulse-echo and (b) continuous-wave ultrasound  techniques.
Here, CT means cross talk.
} 
\label{PECW}
\end{figure}

To solve the problem with the pulse-duration time, we employ the continuous-wave excitation approach instead of the PE method.
The concepts of the PE and CW techniques are represented in Fig.~\ref{PECW}.
The sample with two parallel surfaces is sandwiched between two piezoelectric transducers, where one is used for the ultrasound excitation and the other one is for the detection.
The difference of the two techniques is the excitation waveform.
In the PE technique, an ultrasound-wave pulse with the pulse duration of $\sim0.5$~$\mu$s is excited.
This pulse propagates through the crystal and reflects off the surface.
Each time the ultrasound arrives at the detection transducer, the ultrasound-echo signal is detected.
Here, the echo signals for one-way travel (the so-called 0th echo), three-ways travel (1st echo), and so on are separated in time.
In this case, the influence of the acoustic interference between the incident and reflected waves can be eliminated.
Furthermore, the cross-talk (CT) which comes from the electromagnetic excitation is also separated, since it arrives to the receiver earlier than the acoustic-wave signal.
Therefore, with the PE technique, all the acoustic and electromagnetic noises can be separated in time and efficiently eliminated. 
But this approach works well only if the external parameter (for instance, magnetic field) does not change significantly within at least a few microseconds.

On the other hand, with the CW technique, the ultrasound wave is excited in a continuous manner \cite{Laperre92,Bolef71,Luthi81}.
In this case, the detection transducer receives all the acoustic and electromagnetic signals together.
Here, all reflections and CT interfere with each other.
Only when the first transmitted signal (analog to the 0th echo) is dominant, the acoustic properties of the sample can be extracted.
In this case, the main contribution to the acoustic properties comes from the first transmitted signal, though the contributions from further reflection are significantly suppressed.
Because of this complex situation demanding a sophisticated analysis, the CW technique has not been widely used for ultrasound-velocity measurements in recent years.
However, with this technique, we can obtain useful information on sound-velocity changes at ultrasound frequencies of 10 -- 100~MHz, allowing one to combine the CW technique with STC magnets.

For a successful ultrasound measurement with the CW technique, three requirements have to be met.
First, the 0th echo signal needs to be strong enough compared to the CT.
To satisfy this condition, high-efficiency piezoelectric transducers are necessary.
In this study, LiNbO$_3$ thin plates are used as resonant transducers for the efficient electro-mechanical conversion.
Second, the acoustic attenuation needs to be moderately strong to suppress the higher-echo signals.
Typically, this condition is satisfied close to magnetic-field-induced phase transitions where the acoustic attenuation increases due to the fluctuating  order parameter.
Third, the ultrasound propagation time $\tau$ needs to be reasonably small.
If $\tau$ is too large compared to the magnetic-field variation time, only the averaged ultrasound velocity with changing magnetic field
during $\tau$ is detected.
For reasonably well-defined field value, $\tau$ should be less than 0.5 $\mu$s.
It means that the sample length needs to be reduced for these ultrasound experiments.
However, the measurement accuracy of the sound velocity is proportional to the sample length.
Therefore, the sample length has to be optimized to achieve $\tau \sim 0.3$~$\mu$s.
In this way, the optimal sample length is around 1 mm (0.5 mm) for longitudinal (transverse) acoustic modes.

\subsection{Experimental setup}

\begin{figure}[tb]
\centering
\includegraphics[width=0.99\linewidth]{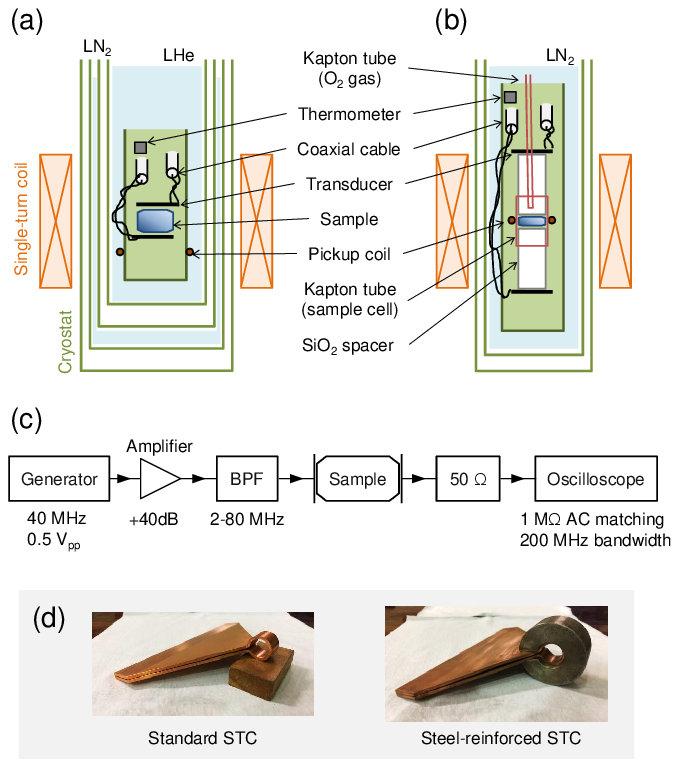}
\caption{
Schematic experimental setting with the sample environment for (a) solid samples and for (b) liquid oxygen.
(c) Block diagram of the electrical circuit. Typical parameters are denoted below the blocks.
BPF is the band-pass filter.
(d) Pictures of the standard and steel-reinforced STCs.} 
\label{experiment}
\end{figure}

Taking into account the above considerations, we designed an experimental setup, allowing us to overcome these and related problems.
The sample with two parallel surfaces was sandwiched by the LiNbO$_3$ transducers to generate and detect ultrasound waves [Fig.~\ref{experiment}(a)].
In this study, Y-36$^\circ$-cut LiNbO$_3$ transducers were used for generating longitudinal acoustic modes.
The transducers were bonded to parallel surfaces of the sample by Thiokol LP31.
The coaxial cable and transducer were connected by the 60 $\mu$m copper twisted-pair wires and the 40 $\mu$m gold wires with silver paint.
The amount of silver paint should be as little as possible to reduce electromotive forces.
In this paper, we present results for two magnetic compounds, MnCr$_2$S$_4$ (cubic, space  group $Fd\overline{3}m$ \cite{Tsurkan17}) and the natural mineral green dioptase Cu$_6$[Si$_6$O$_{18}$] $\cdot$ 6H$_2$O (trigonal, $R\overline{3}$ \cite{PodlesnyakPRB16}) by using the setup as shown in Fig. ~\ref{experiment}(a).
In both samples, the longitudinal acoustic modes were measured in the configurations, 
${\bf k}\parallel{\bf u}\parallel{\bf B}\parallel[111]$
 for MnCr$_2$S$_4$ [($c_{11}+2c_{12}+4c_{44})/3$ mode], and 
${\bf k}\parallel{\bf u}\parallel{\bf B}\parallel[001]$ 
for Cu$_6$[Si$_6$O$_{18}$] $\cdot$ 6H$_2$O ($c_{33}$ mode), respectively.
Here, ${\bf k}$(${\bf u}$) is the propagation (displacement) vector of the acoustic waves.
The sample length, propagation time, and sound velocity at 4~K were 0.96~mm, 0.18~$\mu$s, and 5.3~km/s for MnCr$_2$S$_4$, and 1.37~mm, 0.25~$\mu$s, and 5.5~km/s for Cu$_6$[Si$_6$O$_{18}$] $\cdot$ 6H$_2$O, respectively.

For the case of liquid oxygen, we built a special sample cell sketched in Fig.~\ref{experiment}(b). 
The sample cell was made of two SiO$_2$ glass spacers and Kapton tubes.
These parts were bonded by the low temperature glue, Nitofix SK-229.
The gap between the two SiO$_2$ spacers was filled by high-purity oxygen gas through the thin Kapton tube.
By cooling the sample space, oxygen was condensed from bottom to top of the cell.
Pressure inside the cell was kept at the saturated vapor pressure of liquid oxygen at 77 K.
The transducers were attached to the surface of the SiO$_2$ spacers which acted as acoustic delay lines.
Our experiments revealed that the sound velocity of SiO$_2$ glass does not change up to 100~T with the accuracy of $\Delta v/v < 5 \times 10^{-4}$.
Therefore, the whole change of the sound velocity can be only attributed to liquid oxygen.
Typical lengths of the delay line and the gap thickness were 10 and 0.43 mm ($\tau$ were 1.73 and 0.43~$\mu$s), respectively.
The advantages of such a design are the following.
First, oxygen can be locally condensed in the rigid configuration, which enables to obtain reproducible results in repeated measurements with the same setup.
Second, the transducers are located outside of the STC.
In this case, the risk of the transducer damage is significantly reduced.
Third, with the acoustic delay line, we can discard the data analysis at the beginning of the field pulse, where the electromagnetic noise is the strongest.
Thus, this setup has a potential to realize better signal-to-noise ratio, although the signal amplitude might be reduced due to the delay lines.

The block diagram of the electrical circuit with typical experimental parameters is shown in Fig.~\ref{experiment}(c). 
The transducers were continuously excited during the field generation with a typical peak-to-peak voltage of 50 V.
The excitation frequency was chosen to accord with the fundamental frequency of the transducers of about 40~MHz, which is away from the noise spectrum of the STC system [Fig. \ref{field}(b) inset].
To protect the electrical devices from the huge noise and inductive voltage of the STC system, a band-pass filter (BPF) with the 2 -- 80 MHz transmission band was used.
On detection side, no BPF was used to avoid additional signal reflections.
Note, that the resonant transducers also worked as BPF, allowing the signal transmission only near the resonance frequencies.
Ultrasound signal was recorded by a digital oscilloscope (LeCroy HDO6054) with a sampling rate of 2.5~GHz.
The signal was recorded with the setting of AC 1~M$\Omega$ matching and 200 MHz bandwidth.
Here, an external matching resistance of 50~$\Omega$ was used to protect the oscilloscope.

In our experiments, magnetic fields were generated by a vertical-type STC magnet, designed and installed at the Institute for Solid State Physics, Japan \cite{Miura03}.
The capacitance of the energy supply and maximum charging voltage were 132~$\mu$F and 40 kV, respectively.
The standard STC made of thin copper plate was destroyed in each magnetic-field pulse.
For repeating test measurements, we also used a steel-reinforced STC [Fig. \ref{experiment}(d)] to generate magnetic fields up to 30~T without coil destruction.
In this study, the ultrasound experiments on MnCr$_2$S$_4$ were performed in the reinforced STC magnet.

Magnetic field was measured by the calibrated pickup coil located near the sample.
The recorded pickup voltage proportional to $dB/dt(t)$ was numerically integrated to obtain the magnetic field $B(t)$ (Fig.~1).
Temperature was monitored by a calibrated RuO$_2$ chip resistor installed near the sample.
For cryogenic conditions, a bath cryostat made of epoxy glass was used \cite{12JPSJ_Takeyama}.
The sample and thermometer were immersed in liquid $^4$He (LHe) and cooled down to 4.2 K.
For the case of liquid oxygen, the system was cooled down to 77 K by using liquid nitrogen (LN$_2$).

\section{Results and discussion}
\subsection{MnCr$_\mathbf{2}$S$_\mathbf{4}$}

\begin{figure}[tb]
\centering
\includegraphics[width=0.92\linewidth]{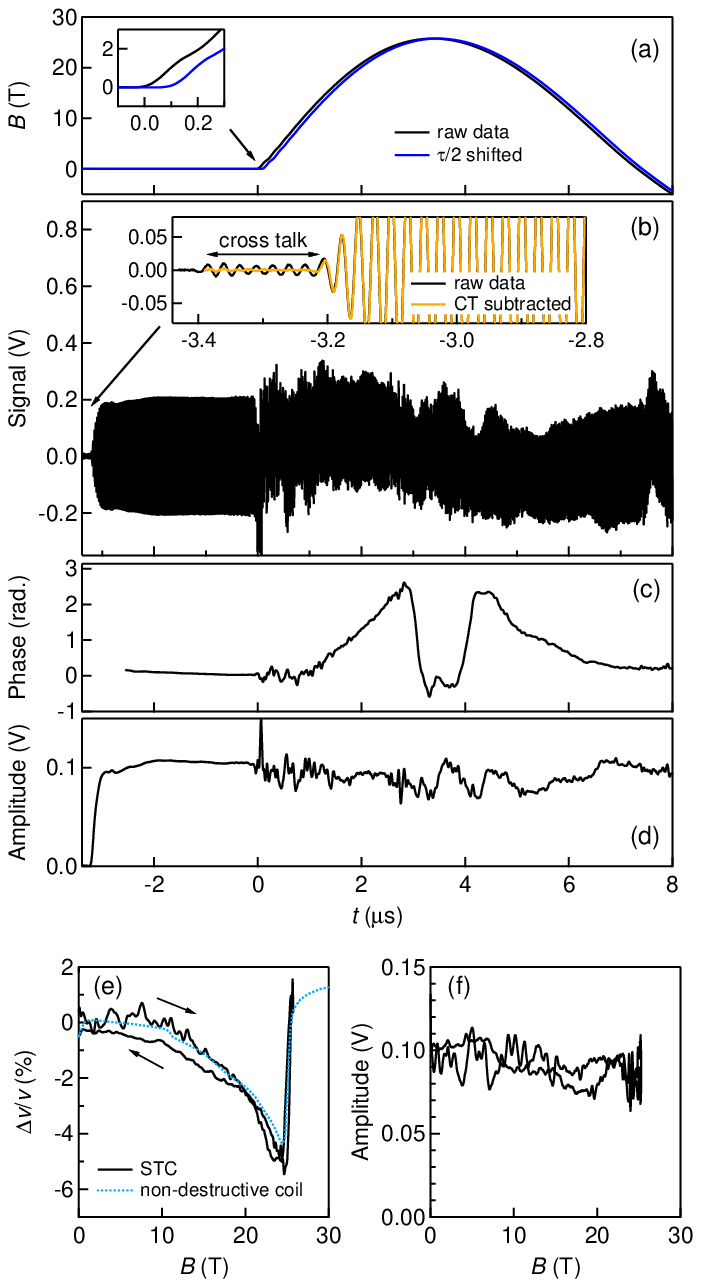}
\caption{
Summarized results of MnCr$_2$S$_4$ obtained by using the steel-reinforced STC.
The measurement temperature was 4.2 K, and the experimental geometry was ${\bf k}\parallel{\bf u}\parallel{\bf B}\parallel[111]$.
(a) Magnetic field waveform. The blue curve is shifted by 92 ns, which corresponds to the half of the propagation time $\tau$. 
(b) Detected signal by the oscilloscope. The inset enlarges the beginning of the sound excitation.
(c) Phase and (d) amplitude of the ultrasound signal obtained by numerical lock-in analysis of the detected signal.
(e) Relative change of the sound velocity and (f) amplitude as a function of magnetic fields.
Results obtained in a field up (down) sweep are plotted by solid (dashed) curves.
The result obtained in a non-destructive coil \cite{Tsurkan17} (within 150 ms magnet pulse) is plotted by the cyan dotted curve for comparison.} 
\label{MnCr2S4}
\end{figure}

First, we present ultrasound results on MnCr$_2$S$_4$ obtained in the steel-reinforced STC.
Figures \ref{MnCr2S4}(a) and \ref{MnCr2S4}(b) show the magnetic-field waveform and the raw data of the acoustic signal, respectively.
The ultrasound frequency is $f=39.6$~MHz.
The excitation starts at $t=-3.4$~$\mu$s, which is observed as a cross-talk (CT) signal.
The ultrasound signal is detected at $\tau \sim 0.18$~$\mu$s after the excitation.
The signal to CT ratio is around 20, which appears to be sufficient to extract the information on the sound velocity and amplitude.
Since the CT is continuously detected with the same phase and amplitude during the excitation, the effect is eliminated by subtracting the sinusoidal wave as in the inset of Fig. \ref{MnCr2S4}(b).
Subsequently, the analysis is done for the data after subtracting the CT parasitic signal.
The obtained signal is analyzed by a digital lock-in analysis.
Computer-generated in-phase and quadrature-phase reference signals are multiplied with the recorded ultrasound signal, and corresponding amplitudes ($I$ and $Q$) are obtained after a digital low-pass filtering.
The ultrasound phase ($\Phi$ from 0 to 2$\pi$) and amplitude $A$ are obtained by calculating $\Phi = \mathrm{arctan}(Q/I)$ and $A=(I^2 + Q^2)^{0.5}$ as shown in Figs. \ref{MnCr2S4}(c) and \ref{MnCr2S4}(d).
A similar procedure is used in the analog data analysis as described in Ref.~[\onlinecite{Luthi05}].
The field-induced phase transition in MnCr$_2$S$_4$, where the magnetic structure changes from canted to collinear type at 25 T \cite{Tsurkan17,Miyata20}, is clearly observed as a change of the ultrasound phase.

The relative change of the sound velocity, $\Delta v/v$, is proportional to the relative change of the phase, $\Delta \Phi/\Phi$ at constant frequency \cite{Luthi05}.
Here, the change of the sample length is usually negligible and ignored.
Thus, $\Delta v/v$ is obtained as,
\begin{equation}
\Delta v/v = -\frac{\Delta \Phi/\Phi}{2\pi f \tau}.
\end{equation}
The sound velocity and amplitude are plotted as a function of magnetic field in Figs. \ref{MnCr2S4}(e) and \ref{MnCr2S4}(f), respectively.
Here, the magnetic-field waveform is delayed by $\tau/2 = 92$~ns [the blue curve in Fig. \ref{MnCr2S4}(a)] because the sound-velocity change is detected with the delay of the propagation time.
Namely, the obtained sound velocity is the averaged value with the time window of $\tau$ and with the delay of $\tau/2$.
As expected, $\Delta v/v$ as a function of $B$ becomes almost reversible by including the effect of the response delay $\tau/2$.
The overall features of $\Delta v/v$, the softening toward 25 T and the clear anomaly at the phase transition, are consistent with the result obtained by using a non-destructive magnet with $10^4$ times longer magnetic field pulse [Fig. \ref{MnCr2S4}(e)] \cite{Tsurkan17}.
The small hysteresis might be due to the magnetocaloric effect in MnCr$_2$S$_4$ \cite{Yamamoto21} because the STC experiments were performed under the quasi-adiabatic conditions.
Therefore, the ultrasound CW technique can be successfully used even with the STC system with pulse duration of a few microseconds.

With the same installation, we expected, the ultrasound measurement could be performed up to 100~T by using the standard STC.
However, we found that the signal-to-noise ratio becomes significantly worse when using the semi-destructive STC magnets with the charging voltage of 40 kV.
This effect might be due to the multiferroic properties of MnCr$_2$S$_4$ \cite{Ruff19}.
In the STC experiment, a huge electromotive force is induced at the beginning of the pulse.
Such voltage might induce structural distortions (piezoelectric effect) and result in resonant oscillations of the single crystal, with the decay time much longer than the pulse duration.
We encountered similar difficulties with two other multiferroic materials, CdCr$_2$O$_4$ \cite{Rossi20} and Fe$_2$Mo$_3$O$_8$ \cite{Wang15,Kurumaji15} as well.
Therefore, the ultrasound measurements for multiferroic crystals appear to be challenging with this setup.
In contrast, in Cu$_6$[Si$_6$O$_{18}$] $\cdot$ 6H$_2$O which is not multiferroic, the ultrasound measurement can be performed up to $\sim 100$~T as discussed in the next section.

\subsection{Cu$_\mathbf{6}$[Si$_\mathbf{6}$O$_\mathbf{18}$] $\cdot$ 6H$_\mathbf{2}$O}
\begin{figure}[!tbh]
\centering
\includegraphics[width=0.92\linewidth]{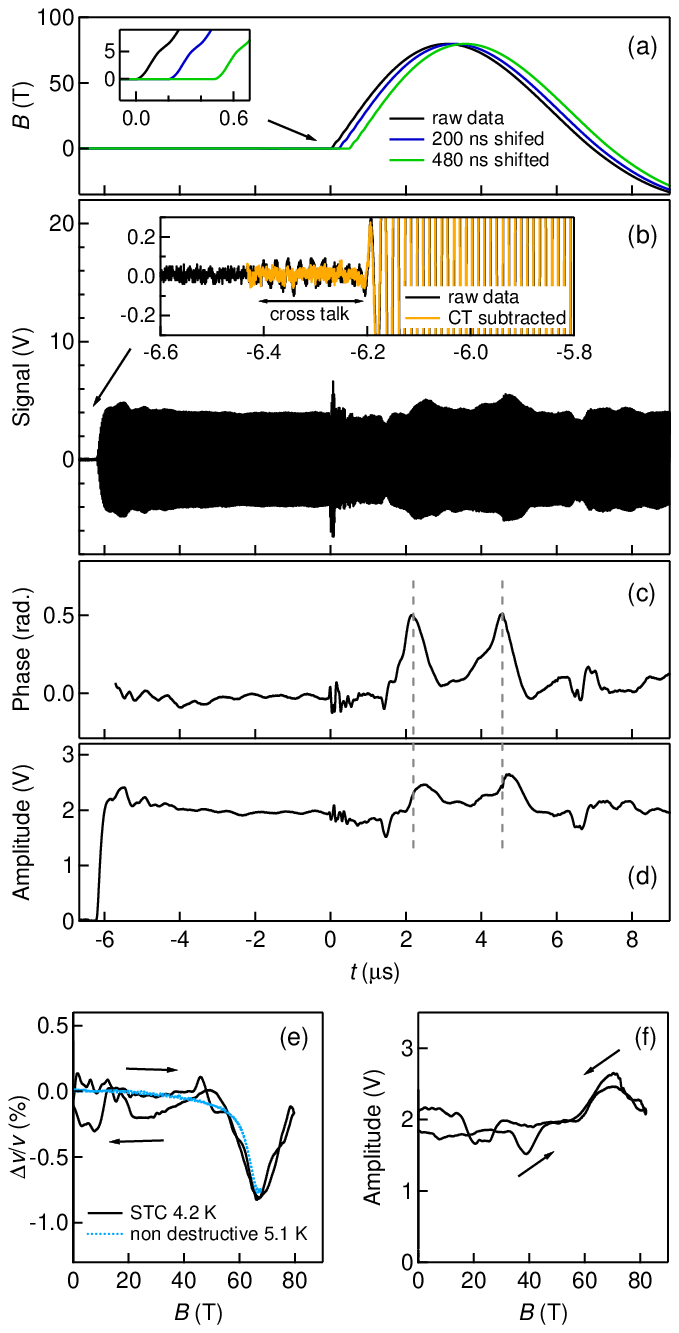}
\caption{
Summarized results for Cu$_6$[Si$_6$O$_{18}$] $\cdot$ 6H$_2$O by using the standard STC magnet.
The measurement temperature was 4.2 K, and the experimental geometry was ${\bf k}\parallel{\bf u}\parallel{\bf B}\parallel[001]$.
(a) Magnetic field waveform. The blue curve is shifted by 200 ns.
(b) Detected signal by the oscilloscope. The inset enlarges the beginning of the excitation.
(c) Phase and (d) amplitude of the ultrasound signal.
The gray dashed lines indicate the anomalies detected in phase of the ultrasound signal.
(e) Relative change of the sound velocity and (f) amplitude as a function of magnetic fields.
The sound velocity and amplitude are plotted as a function of 200-ns-shifted field profile.
Results obtained in a field up (down) sweep are plotted by solid (dashed) curves.
The results obtained by a non-destructive magnet with 150 ms pulse duration at an ultrasound frequency of 104 MHz are plotted by the cyan dotted curves for comparison.} 
\label{GD}
\end{figure}

Here, we present ultrasound results for the natural mineral green dioptase Cu$_6$[Si$_6$O$_{18}$] $\cdot$ 6H$_2$O obtained with the standard STC.
By using a 16-mm-diameter STC with 36 kV charging voltage, the magnetic field of up to 80~T was generated [Fig.~\ref{GD}(a)].
The measurement can easily be extended up to 120 T by using a 12-mm-diameter STC.
In this research, we preferred the moderate field-sweep rate to focus on the phase transition expected around 70 T.
The raw data of the acoustic signal are shown in Fig.~\ref{GD}(b).
The ultrasound frequency is $f=42.0$~MHz.
Thanks to the relatively low sound attenuation in the Cu$_6$[Si$_6$O$_{18}$] $\cdot$ 6H$_2$O single crystal, we obtained rather strong signal compared to the one presented in section A (Fig.~\ref{MnCr2S4}).
The signal to CT ratio was around 50, and the signal was much larger than the starting noise at $t=0$~$\mu$s.
By using the same analytical procedure as in the case of MnCr$_2$S$_4$, the phase and amplitude of the acoustic signal are obtained [Figs.~\ref{GD}(c) and (d)].
Two clear anomalies, indicated by the gray dashed lines for the field up and down sweeps, are observed both in the phase and amplitude of the ultrasound signal.
Although the phase and amplitude typically behave differently at a phase transition, it seems that the anomalies in amplitude appear slightly later than those in phase.

Figures \ref{GD}(e) and \ref{GD}(f) show the $\Delta v/v$ and amplitude of the ultrasound signal as a function of $B$, respectively.
The clear anomaly is observed around 70 T, indicating the magnetic phase transition in Cu$_6$[Si$_6$O$_{18}$] $\cdot$ 6H$_2$O.
The mechanism of this transition is discussed elsewhere \cite{prep}.
Because no any significant hysteresis has been observed in ultrasound experiments in the non-destructive pulsed magnet, the 200~ns delay [Fig.~\ref{GD}(a)] is chosen to minimize the hysteresis loop of $\Delta v/v$ in a field region from 50 to 80~T. 
We find that the delay $\tau/2 \sim 130$~ns is not sufficient for this purpose.	
Note that the magnetocaloric effect in the green dioptase measured up to 60 T is too small to explain the observed hysteresis \cite{prep}.

Regarding the amplitude, even 200-ns delay cannot minimize the observed hysteresis.
Besides, the ultrasound amplitude increases near the phase transition at 70~T [Fig.~\ref{GD}(f)], in contrast to a practically constant level observed in the PE experiments.
The inconsistency of the amplitude change is considered to be due to the complex interference of the acoustic waves inside the crystal.
In our experimental setting, the transducers covered the whole polished surfaces of the crystal.
However, the side surfaces of the crystal had irregular shape and were not polished to keep the sample as large as possible.
In such case, acoustic waves are irregularly reflected or scattered at side surfaces, resulting in the complex interference which is not realistic to decompose by the numerical analysis.
By preparing a rectangular-shaped crystal, the effect of interference might be accounted for by numerical simulations in the future.
In this case, the contribution of the reflected waves could be estimated by measuring the frequency dependence of the signal amplitude, since the amplitude of the standing wave strongly depends on the wave phase at the boundaries.
Nevertheless, in our case, the first transmitted signal appears to be still dominant because the obtained $\Delta v/v$ versus $B$ curve well reproduces the result obtained with the PE technique [Fig.~\ref{GD}(e)].
The obtained $\Delta v/v$ resolution is of the order of 10$^{-3}$, which is sufficient for studying most field-induced phase transitions.

Here, we discuss reasons of the difference in delays for the phase and amplitude [Fig. \ref{GD}(c) and \ref{GD}(d)].
These different response times are difficult to explain by the acoustic interference between the transmitted and reflected waves.
One possible reason is the nonlinear phonon dispersion near the field-induced phase transition.
At the magnetic phase transition, the phonon dispersion near the $\Gamma$ point in the Brillouin zone might be deformed due to the hybridization with a magnon soft mode \cite{Kittel58,Nomura_PMCE19}.
As a result, the phase velocity ($\Delta \omega / \Delta k$) and group velocity ($d\omega / dk$) of the ultrasound become different.
Here $\omega$ is the angular frequency and $k$ is the wavenumber.
These difference results in a time mismatch of the characteristic phase and amplitude anomalies close to the phase transition, since the amplitude and phase signals propagate with group and phase velocities, respectively.
Therefore, the response-time difference for the ultrasound phase and amplitude might indicate the nonlinear acoustic phonon dispersion.
In addition, a spin-lattice decoupling might play an important role in the microsecond time scale. 
If the field-variation time is shorter or comparable to the lattice relaxation time, the lattice response can be delayed even if the spins follow the magnetic field. 
Such response delay has been observed in organic compounds even in slow changing continuous magnetic fields \cite{Isono18}.
In the STC experiments, the partial spin-lattice decoupling might result in the larger response delay than the expected value.

\subsection{Liquid oxygen}

\begin{figure}[!tbh]
\centering
\includegraphics[width=0.92\linewidth]{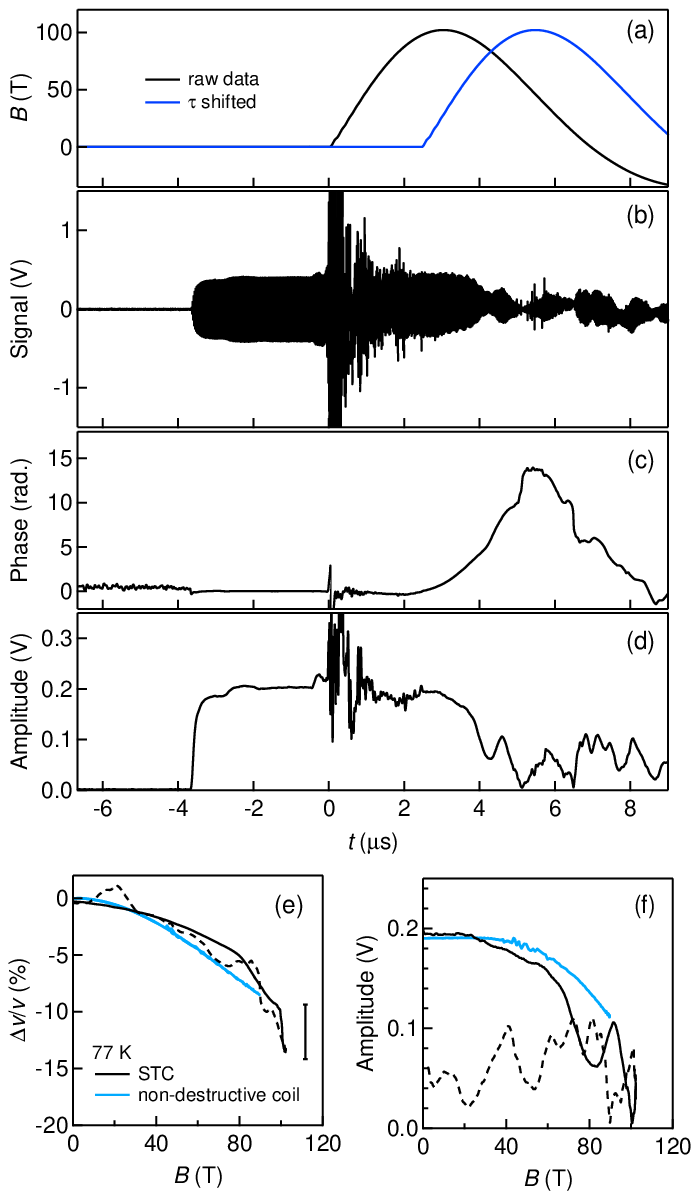}
\caption{
Summarized results for liquid oxygen by using the standard STC magnet.
The measurement temperature was 77 K, and the experimental geometry was ${\bf k}\parallel{\bf u}\parallel{\bf B}$.
(a) Magnetic field waveform. The blue curve is shifted by 2.4 $\mu$s, which roughly corresponds to the propagation time in the acoustic delay line and sample.
(b) Detected signal by the oscilloscope. 
(c) Phase and (d) amplitude of the ultrasound signal.
(e) Relative change of the sound velocity and (f) amplitude as a function of the shifted magnetic field (see text for details).
Results obtained in a field up (down) sweep are plotted by solid (dashed) curves.
Cyan curves show the results obtained in a non-destructive magnet \cite{Nomura19}.} 
\label{LO2}
\end{figure}

Finally, we present results of liquid oxygen obtained with the acoustic delay lines as shown in Fig.~\ref{experiment}(b).
By using a 14-mm-diameter STC with 40 kV charging voltage, the magnetic field up to 103~T was generated [Fig.~\ref{LO2}(a)].
Because of the acoustic delay lines, the field profile has to be shifted by 2.4~$\mu$s to minimize the hysteresis loop.
This time delay roughly corresponds to the total delay expected for one delay line (1.7 $\mu$s) plus half of the sample length (0.2 $\mu$s).
The deviation from the expected time delay might be due to the acoustic interference or error for estimating the sample length, which is measured at room temperature.
Figure ~\ref{LO2}(b) shows the detected signal as a function of time.
The ultrasound frequency is 38.8~MHz.
Note that the CT is rather small in this experiment because the distance between two transducers is much larger than for the configuration shown in Fig. \ref{experiment}(a).
Thus, the analysis can be done even without subtracting the CT contribution.
By the numerical lock-in analysis, the phase and amplitude of the signal are obtained as in Figs.~\ref{LO2}(c) and \ref{LO2}(d).
Note, that the phase is unwrapped at $2\pi$ and $4\pi$.
Both the phase and amplitude become unstable at $\sim6$~$\mu$s after the pulse begins, probably due to the mechanical vibration of liquid oxygen.
Since the liquid oxygen is a paramagnetic liquid, it is attracted to the center of the magnet in applied magnetic field.
Besides, the volume increase of liquid oxygen in magnetic field \cite{Uyeda87} could also result in the vibration or convection.
Within the $\mu$s time scale, the motion of liquid oxygen cannot be damped.
Therefore, the results obtained in the field-up sweep is more reliable than in the field-down sweep.

The obtained $\Delta v/v$ and ultrasound amplitude are plotted as a function of magnetic field in Figs. \ref{LO2}(e) and \ref{LO2}(f), respectively.
The results of the field-up (field-down) sweeps are shown by solid (dashed) curves.
The results obtained in the non-destructive pulsed magnet with much longer pulse duration are shown by the cyan curves for comparison \cite{Nomura19}.
Note that the amplitude, $A$, is normalized to the corresponding sample length according to the relation, $A=A_0 \mathrm{e}^{-\alpha d}$, where $\alpha$ is the acoustic attenuation coefficient, and $d$ is the distance.
The relative changes of the sound velocity and amplitude agree well for the data obtained with the different time scales by a factor of $10^4$.
The small oscillations observed in the STC for the field-up sweep are considered to be due to the interference between the incident and reflected ultrasound waves with the destructive and constructive phase shifts.
Note that the magnetocaloric effect in liquid oxygen is less than 1 K temperature increase at 50 T \cite{Nomura17MCE}, which is negligibly small to explain the observed change of $\Delta v/v$.

By using the non-destructive pulsed magnet, the softening of 8~\% and the continuous increase of the acoustic attenuation were observed up to 90 T, which might be a precursor of a liquid-liquid transition (LLT) in oxygen \cite{Nomura19}.
In this study, we extended the field range of measurements up to 100 T.
By using the developed technique, high-field ultrasound measurements are feasible even beyond 100 T to investigate various phase transitions including the LLT of oxygen.

\section{Conclusion}
We have demonstrated the ultrasound technique combined with the STC is feasible in magnetic fields beyond 100 T within microsecond time scale.
By using the STCs with the inner magnet bore of 8~mm, such experiments can be extended up to 200 T.
To perform the ultrasound experiments within this timescale, we employed the CW ultrasound method.
The developed technique was successfully applied to several magnetic systems,  MnCr$_2$S$_4$, Cu$_6$[Si$_6$O$_{18}$] $\cdot$ 6H$_2$O, and liquid oxygen, where a softening of the acoustic modes related to various phase transitions occurs at high fields.
There is a fair agreement between the STC results and the results obtained by the conventional PE technique in a non-destructive pulsed magnet providing $10^4$ longer time scale.
For the case of single-crystal samples, the resolution of $\Delta v/v$ in the STC is of the order of 10$^{-3}$.

\section*{Acknowledgments}
This work was partly supported by the JSPS KAKENHI Grants-In-Aid for Scientific Research (No. 19K23421, No. 20K14403) and JSPS Bilateral Joint Research Projects (JPJSBP120193507).
We acknowledge support of the HLD at HZDR, member of the European Magnetic Field Laboratory (EMFL), the BMBF via DAAD (project-id 57457940), and the DFG through ZV 6/2-2, SFB 1143, Transregional Research Collaboration TRR 80 (Augsburg, Munich, and Stuttgart), and the W\"urzburg-Dresden Cluster of Excellence on Complexity and Topology in Quantum Matter--$ct.qmat$ (EXC 2147, project No. 390858490).
We also acknowledge the project ANCD 20.80009.5007.19 (Moldova).
We would like to thank A. Podlesnyak for providing a green dioptase sample.

\section*{Data availability}
The data that support the findings of this study are available from the corresponding author upon request.

%\clearpage

\end{document}